\documentclass[a4paper,twocolumn,superscriptaddress,11pt,longbibliography]{quantumarticle}

\usepackage[utf8]{inputenc}
\usepackage[english]{babel}
\usepackage[T1]{fontenc}

\usepackage{amsmath}
\usepackage{amsfonts}
\usepackage{amssymb}
\usepackage{amsthm}
\usepackage{hyperref}
\usepackage[numbers,sort&compress]{natbib}
\usepackage{tikz}
\usepackage{lipsum}
\usepackage[normalem]{ulem}
\usepackage{exscale}
\usepackage{bbm}
\usepackage{graphicx}
\usepackage{latexsym}
\usepackage{enumerate}
\usepackage{bbold}
\usepackage{color}
\usepackage{nicefrac}
\usepackage{mathtools}
\usepackage{chngcntr}
\usepackage{tcolorbox}
\usepackage{wrapfig}
\usetikzlibrary{decorations.pathreplacing}
\usetikzlibrary{graphs}
\usepackage{tikz}
\usepackage{pgfplots}
\usepackage{array, threeparttable, booktabs}

\usepackage{subcaption}
\newcommand{\ket}[1]{\left|#1\right\rangle}

\theoremstyle{definition}

\theoremstyle{remark}

\definecolor{quantumviolet}{HTML}{53257F} 
\definecolor{quantumgray}{HTML}{555555} 
\definecolor{mygray}{gray}{0.95} 

\newtcolorbox[auto counter,number within=section]{boxfigure}[2][]{%
colback=mygray,colframe=quantumviolet,fonttitle=\bfseries,width=\textwidth,float*=ht,lower separated=false, halign=justify,title=Box~\thetcbcounter: #2,#1}

\usepackage{listings} 
\definecolor{commentsColor}{rgb}{0.497495, 0.497587, 0.497464}
\definecolor{keywordsColor}{rgb}{0.000000, 0.000000, 0.635294}
\definecolor{stringColor}{rgb}{0.558215, 0.000000, 0.135316}

\usepackage{tikz}
\usetikzlibrary{quantikz}
\usepackage{tikz-cd}
\tikzcdset{scale cd/.style={every label/.append style={scale=#1},
		cells={nodes={scale=#1}}}}
\usepackage{physics}
\usepackage[outline]{contour} 
\usetikzlibrary{intersections}
\usetikzlibrary{decorations.markings}
\usetikzlibrary{angles,quotes} 
\usetikzlibrary{calc}
\usetikzlibrary{3d}

\tikzset{>=latex} 
\colorlet{myred}{red!85!black}
\colorlet{myblue}{blue!80!black}
\colorlet{mycyan}{cyan!80!black}
\colorlet{mygreen}{green!70!black}
\colorlet{myorange}{orange!90!black!80}
\colorlet{mypurple}{red!50!blue!90!black!80}
\colorlet{mydarkred}{myred!80!black}
\colorlet{mydarkblue}{myblue!80!black}
\tikzstyle{xline}=[myblue,thick]

\tikzstyle{myarr}=[myblue!50,-{Latex[length=3,width=2]}]

\usepackage{lipsum}
\usepackage{adjustbox}
\usepackage{longtable}
\usepackage{colortbl}
\usetikzlibrary{arrows}
\usepackage{makeidx}\makeindex
\tikzset{
	operator/.append style={fill=purple!20},
	my label/.append style={above right,xshift=0.3cm},
	phase label/.append style={label position=above}
}

\newcommand*{\gateStyle}[1]{{\textsf{\itshape #1}}}
\newcommand*{\hGate}{\gateStyle{H}}
\newcommand*{\uGate}{\gateStyle{U}}
\newcommand*{\xGate}{\gateStyle{X}}

\usepackage{silence}
\newcommand*{\circuitH}{\gate[style={fill=teal!20},label style=black]{\textnormal{\hGate{}}}}
\newcommand*{\circuitU}{\gate[style={fill=arsenic!20},label style=black]{\textnormal{\uGate{}}}}
\newcommand*{\circuitX}{\gate[style={fill=blue!20},label style=black]{\textnormal{\xGate}}}

\definecolor{arsenic}{rgb}{0.23, 0.27, 0.29}
\definecolor{fluxcolor}{RGB}{204, 217, 255}
\definecolor{uwavecolor}{RGB}{244, 220, 222}
\definecolor{FandUwavecolor}{RGB}{231,244,224}
\definecolor{cavitycolor}{RGB}{232, 200, 244}
\definecolor{orangeb}{rgb}{0.99,0.78,0.07}
\definecolor{orangebdark}{rgb}{0.99,0.78,0.17}
\definecolor{livingcoral}{HTML}{FA7268}			
\definecolor{ultraviolet}{HTML}{5F4B8B}			
\definecolor{greenery}{HTML}{88B04B}			
\definecolor{radiantorchid}{HTML}{AD5E99}		
\definecolor{tangerinetango}{HTML}{DD4124}		

\DefineNamedColor{named}{Salmon}        {cmyk}{0,0.53,0.38,0}
\DefineNamedColor{named}{CarnationPink} {cmyk}{0,0.63,0,0}

\usepackage{float}

\usepackage{imakeidx}
\makeindex[columns=2, title=Alphabetical Index,  options= -s alpha.ist]

\definecolor{commentsColor}{rgb}{0.497495, 0.497587, 0.497464}
\definecolor{keywordsColor}{rgb}{0.000000, 0.000000, 0.735294}
\definecolor{stringColor}{rgb}{0.558215, 0.000000, 0.135316}
\definecolor{carrotorange}{rgb}{0.93, 0.57, 0.13}
\usepackage{tabularx}
\usepackage{ltablex}
\usepackage{caption}

\DeclareCaptionLabelFormat{boxed}{%
	\kern0.01em{\color{carrotorange}\rule{0.73em}{0.73em}}%
	\hspace*{0.47em}\bothIfFirst{#1}{~}#2}
\tikzstyle{input}=[draw,fill=red!50]
\tikzstyle{conv}=[draw,fill=black!20]
\tikzstyle{max}=[draw,dashed,fill=black!10]
\tikzstyle{dropout}=[draw,dashed,fill=blue!10]
\tikzstyle{fc}=[draw,fill=green!10]
\tikzstyle{output}=[draw,fill=red!50]

\usepackage[europeancurrents, europeanvoltages, americanresistors, cuteinductors, americanports, nosiunitx, noarrowmos]{circuitikz}

\definecolor{BlueLUH}{cmyk}{1.0,0.7,0,0}
\colorlet{LightBlue}{BlueLUH!20!white}
\colorlet{DarkBlue}{BlueLUH!80!black!20}
\colorlet{PinKish}{red!60}

\usepackage{tikz}
\usetikzlibrary{arrows,automata}

\usepackage{enumitem}
\setlist{nosep,leftmargin=\leftmargin/2}

\usepackage[utf8]{inputenc}
\usepackage[T1]{fontenc}
\usepackage[normalem]{ulem}

\newcommand{\quotebox}[1]{\begin{center}\fcolorbox{white}{blue!5!gray!5}{\begin{minipage}{0.9\linewidth}\vspace{5pt}\center\begin{minipage}{0.8\linewidth}{\space\Huge``}{#1}{\hspace{1em}\break\null\Huge\hfill''}\end{minipage}\smallbreak\end{minipage}}\end{center}}

\usepackage{csquotes}

\usepackage{caption}
\begin{document}

\title{Using the Julia framework to teach quantum entanglement.}
\index{Entanglement}

\author{\raisebox{-0.2em}{\textcolor{FandUwavecolor}{\rule{0.8em}{0.8em}}} Shlomo Kashani \textsuperscript{$\Phi$}}
\affiliation{Graduate School (Applied Physics), Johns Hopkins University, Maryland U.S.A.: \url{skashan2@jh.edu}}

\author{\raisebox{-0.2em}{\textcolor{fluxcolor}{\rule{0.8em}{0.8em}}} Prof. David Zaret \textsuperscript{$\Psi$}}
\affiliation{Principal Professional Staff, JHU Applied Physics Laboratory, Maryland, U.S.A: \url{david.zaret@jhuapl.edu}}


\maketitle

\onecolumn
\begin{abstract}
\index{Hidden-variable models}

%

Entanglement, a phenomenon that has puzzled scientists since its discovery, has been extensively studied by many researchers through both theoretical and experimental means \cite{Frauchiger2018, Aharonov2002, Ghirardi2005, Greenberger7, Mermin1998, Levesque2007, woo}. It is a fundamental aspect of both quantum information processing (QIP) and quantum mechanics (QM).

But how can entanglement be most effectively taught  \cite{Zhu2011,Mermin2003, Mykhailova2020} to computer science students compared to applied physics students?. in this educational pursuit, we propose using Yao.jl \cite{Luo2020Yao}, a quantum computing framework written in Julia \cite{kramer2018quantumoptics, julia2015} for teaching entanglement to graduate computer science students attending a quantum computing class \cite{Zaret2022} at Johns Hopkins University.

David Mermin's  \textit{\textbf{ just enough QM} for them to understand and develop algorithms in quantum computation} \cite{Mermin1998, Mermin2003} idea aligns with the purpose of this work. Additionally, the authors of the study \cite{Zhu2011} \textit{Improving students’ understanding of QM via the Stern-Gerlach experiment (SGE)} argue that this experiment should be a key part of any QM education.

Here, we explore the concept of entanglement and it's quantification in various quantum information processing experiments, including one inequality-free \cite{Ghirardi2005, Greenberger7,Aharonov2002} form of Bell's theorem \cite{Mermin1998}: (1) Superposition via the Hadamard (\ref{sub:sup:had}), (2) Bell-state generation (\ref{sub:sup:had1}) and (3) GHZ state generation (\ref{sub:sup:had2}). The utilisation of circuit diagrams \cite{quirk2016} and  code fragments \cite{eigenweb,Gheorghiu2018} is a central theme in this work's philosophy.
\index{Quantum circuit}
\end{abstract}
\index{Entanglement}
\index{Bell's theorem}
\index{SGE}
\index{SGE}
\index{Superposition}
\index{Hadamard}
\index{Yao.jl}
\index{Julia}

\setcounter{tocdepth}{4}
\tableofcontents


\section{Introduction}
\label{sec:introduction}

Here, we provide a comprehensive plan for computer science institutions to effectively teach the fundamental concept of entanglement using the Julia language, while targeting computer science students who may not have a strong background in QM. We present each concept of entanglement from \textbf{\textit{multiple perspectives}}, including the \textbf{\textit{mathematical equations, the corresponding quantum circuit, the Julia code for generating the circuit}}, and, in some cases, the result of simulating the circuit \textbf{\textit{on a classical computer}}. This approach allows students to understand the concept from different perspectives and helps to reinforce their learning.
\index{Quantum circuit}

In the realm of teaching QIS \cite{shoshani2021,nielsen00} and quantum computing, courses often encompass a variety of subjects such as quantum algorithms, cryptography, programming, laboratory work, and hardware design. However, this analysis highlights several key areas of difficulty for students, including difficulties with quantum  formalism and the need for basic quantum programming assignments that focus on entanglement.
\index{QIS}
\index{Quantum information processing}

For teaching engagement, there are several quantum computing libraries, such as ~\citep{reqwire, liquid, ying-foundations, qml, qpl, quantum-lambda-calc, silq,qsharp, quipper, qwire, clairambault20, staton17}, which allow quantum scientists to design and create quantum circuits for use on actual quantum computers. These libraries, which are primarily written in Python \cite{qiskit, bergholm2018pennylane,qsharp}, are developed by companies that aim to run quantum computing algorithms on real hardware rather than classical simulations for educational purposes. However, there are several reasons why it is beneficial to use a classical simulator when teaching graduate students about quantum computing:


\begin{enumerate}
	\item  \raisebox{-0.2em}{\textcolor{cavitycolor}{\rule{0.8em}{0.8em}}} Due to the high error rates of current quantum computers \cite{nielsen_chuang_2010}, research on quantum error correction is a highly active field. However, for the purpose of comparing a student's mathematical derivation to the optimal mathematical results, a classical simulator may be more suitable.
	\item  \raisebox{-0.2em}{\textcolor{cavitycolor}{\rule{0.8em}{0.8em}}} Quantum hardware tests do not always provide immediate measurement for all values, including the full quantum state vector and marginal probabilities of measurements.
	\item  \raisebox{-0.2em}{\textcolor{cavitycolor}{\rule{0.8em}{0.8em}}} While access to actual quantum hardware is not as restricted as it used to be \cite{Mykhailova2020}, it is still not widely accessible in all locations and educational institutions. In contrast, classical simulators can be run on most modern computers.
\end{enumerate}

\subsection{Prior work}
\label{sec::sge}

\subsubsection{Educational programmes}
Universities are offering master's degree programs in quantum computing to meet the increasing demand for skilled professionals in this field. These programs are generally geared towards students who have already completed a bachelor's degree in a STEM field, such as computer science, physics, or engineering, and want to specialize in quantum computing. However, these programs may struggle to accommodate students with diverse technical backgrounds and different levels of knowledge about QM.

Some students may have a strong foundation in computer science but may be unfamiliar with QM, while others may have a strong background in QM but may lack knowledge of computer science. This can make it difficult for universities to design a curriculum that meets the needs of all students and helps them fully understand the complex concepts of quantum computing.

For instance, in the UK, UCL offers a master's programme in Quantum technologies \cite{UCL2023} emphasizing quantum simulation: \\
\textit{"The programme prepares graduates for careers in the emerging quantum technology industries which play an increasingly important role in: secure communication; sensing and metrology; the simulation of other quantum systems; and ultimately in general-purpose quantum computation. Graduates will also be well prepared for research at the highest level in the numerous groups now developing quantum technologies and for work in government laboratories."} Upon reviewing the \href{https://www.ucl.ac.uk/module-catalogue/modules/advanced-quantum-theory-PHAS0069}{curriculum}, it becomes apparent that it may not be suitable for computer science students due to its strong focus on QM.

In Zurich ETHZ offers a master's programme in Quantum Engineering \cite{ethz2023} where it is also apparent that it mainly targets applied physics students:\\
\textit{"A quantum engineer harnesses the laws of QM to provide technological solutions to problems currently unsolvable using classical resources. To tackle this formidable task, on the one hand, a quantum engineer needs to be well versed in \textbf{quantum theory} .... "}

\subsubsection{Recommended quantum-aware  teaching methodologies}
To address these concerns, several authors which we explore here, have proposed various approaches to address this issue and ensure that these programs can effectively educate a diverse group of students including computer science students. In particular, we refer the reader to the following studies.
\begin{enumerate}

	\item  \raisebox{-0.2em}{\textcolor{cavitycolor}{\rule{0.8em}{0.8em}}} In \cite{Asfaw2022}, the authors have developed a comprehensive and innovative curriculum for training \textbf{\textit{"quantum-aware engineers"}} in quantum engineering. The curriculum includes several modules, such as "Classical Information Theory" and "Two-Qubit Gates and Entanglement," which are intended for STEM students, as well as an advanced module called "Hamiltonians and Time Evolution." While this curriculum is well-suited for a full degree program, it cannot be condensed into a single semester-long module on quantum computation.

	\item  \raisebox{-0.2em}{\textcolor{cavitycolor}{\rule{0.8em}{0.8em}}} In the work by Microsoft's team entitled \textit{Teaching QC through a Practical Software-driven Approach: Experience Report} \cite{Mykhailova2020}, the authors also highlight the challenges encountered by computer science undergraduates, whom they were charged with teaching QC, not least among their findings is this:\\
	\textit{"Teach QC through software engineering instead of physics. We approached the course as computer scientists, deliberately \textbf{avoiding physics in the lectures}. We presented the qubits as abstract objects described by vectors instead of quantum mechanical states, and the operations on them as matrix transformations instead of physical processes."} We on the other hand, do think that some, even minimal introduction to the concepts of projections, a measurement apparatus, eigenvalues, and operators is \textbf{essential to the end-to-end understanding of entanglement}.

	\item  \raisebox{-0.2em}{\textcolor{cavitycolor}{\rule{0.8em}{0.8em}}} Other attempts at teaching entanglement were made via the construction of quantum versions of classical games. In \cite{Allan2006}, the authors attempted to teach entanglement through the use of a quantum version of the tic-tac-toe game. They claim that this approach allows for a better understanding of entanglement. They state that the game:\\
	\textit{"offers a way of introducing QM without advanced mathematics"} and conclude that \textit{"Quantum tic-tac-toe illustrates a number of quantum principles including \textbf{states, superposition, collapse, nonlocality, entanglement}, the correspondence principle, interference, and decoherence"}.
	\item  \raisebox{-0.2em}{\textcolor{cavitycolor}{\rule{0.8em}{0.8em}}} And finally, a quote from David Mermin's \textit{From Cbits to Qbits: \textbf{Teaching computer scientists QM}} \cite{Mermin2003} as Mermin is a highly respected pioneer in the field of quantum computing:\\
	\quotebox{A strategy is suggested for teaching mathematically literate students, \textbf{with no background in physics},\textbf{ just enough QM} for them to understand and develop algorithms in quantum computation and quantum information theory\cite[p.\,3]{Mermin2003}}

\end{enumerate}

Following \cite{Mermin2003}, one notable example of an experiment that can be used to teach entanglement is the SG experiment which is presented Chapter 1 of Nielsen's \cite{nielsen_chuang_2010} quantum information processing (Figure 1.22 and equations 1.56 through 1.59) and later on the cascaded SG experiment (Figure 1.23 and equations 1.56 through 1.59). We will now examine the SGE from a quantum computing perspective in more depth.

\index{Entanglement}
\subsection{The important concepts of the SGE}
Before discussing Entanglement, we recommend that at least one or two classes in a quantum computing module be dedicated to the study of the SG experiment \cite{JamesCresser,nielsen_chuang_2010, Montague}. Graduate students with some knowledge of QM should be able to understand the main points covered in this material.
\index{SGE}

There are numerous experimental verifications of QM, but one of the most important and surprising ones is the SG experiment. In it, randomly oriented electrons are shot through a non-uniform electric field whose gradient is oriented in the +z-direction \cite{Zhu2011}. Specifically, a narrow beam of neutral spin-$1/2$ particles is directed along a specific $n$-axis (usually $z,y, or z$) through a S-G ( $z$ axis) apparatus. The apparatus, allows a beam consisting of neutral particles in different spin states to split into different beams \cite{nielsen00}.

\begin{figure}[H]
	\centering
	\includegraphics[totalheight=2cm]{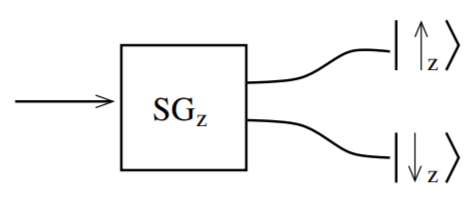}
	\caption{The figure depicts the SG experiment oriented in the z-direction (SGz). Source \cite{babis}.}
	\label{fig::sg01}
\end{figure}
\index{SGE}
\index{S-G apparatus}
\index{Polarized}
\index{Z-axis}
\index{Eigenvalue}

But what is the connection between the SGE and entanglement, and why do we often teach the SGE as a precursor to understanding entanglement? The answer is that the \textbf{\textit{angular momentum of one particle can be entangled with the angular momentum of another particle}}. Therefore, the relationship between the SGE and entanglement is that both involve the concept of angular momentum and its quantization, but the SGE demonstrates this concept in a single particle, while entanglement involves the correlation of angular momentum between two or more particles

Probabilities also play a crucial rule in entangelment  \cite{nielsen_chuang_2010}. Here, the computation of the respective probability amplitude $\langle\phi \mid \psi\rangle$ (also termed a \textbf{projection}), which is a complex number providing the probability amplitude of finding state $|\psi\rangle$ in state $|\phi\rangle$, plays a central role in the experiment. The experiments in Fig. (\ref{fig::sg01}) (\cite{Levesque2007}, \cite{JamesCresser}) reveal that it is possible to send all particles through a Z-axis oriented apparatus and get a $50\% / 50\%$ distribution and subsequently remove one state completely from the system (for instance, by using a blocker), and send it through an X-axis oriented S-G ($x$ axis) apparatus and get $50\% / 50\%$ distribution once again.

Finally, if we take half of these away and put the remaining ones through another S-G ($z$ axis) (e.g. Z-axis oriented apparatus) we end up with a $50\% / 50\%$ distribution once again. This illustrates a fundamental quantum mechanical postulate - \textbf{\textit{the only values that are observed in a measurement are the eigenvalues of the measurement operator}}, where the SG filters are also represented and treated as operators.

For spin-$1/2$ particles, a beam polarized along $z_{+}$ direction is defined as:
\begin{equation}
	\label{eign001}
	z_{+}=\left(\begin{array}{l}
		1 \\
		0
	\end{array}\right),\left(\text { eigenvalue }+\frac{\hbar}{2}\right) ; \quad z_{-}=\left(\begin{array}{l}
		0 \\
		1
	\end{array}\right), \quad\left(\text { eigenvalue }-\frac{\hbar}{2}\right)
\end{equation}

\textit{\textbf{Note the use of the $+\hbar/2$ notation, which is very commonly used in physics but may be unknown to CS students}}. The S-G apparatus measures the projection of $\mathbf{S}$ in a direction of the $z$ axis. There can only be two outcomes, $+\hbar / 2$ and $-\hbar / 2$. These eigenvalues correspond to two distinct eigenstates, which are considered to be orthogonal and are usually written using Dirac notation as $|\uparrow\rangle$ and $|\downarrow\rangle$, or in the context of quantum computing as $|0\rangle$ and $|1\rangle$.
\index{Spin-half}
\index{Eigenvalues}
\index{Eigenstates}
\begin{figure}[H]
	\begin{lstlisting}[caption={},language=C++,]
		oven = (0_ket + 1_ket).normalized();
		resulting_state= measure(oven,Z);
	\end{lstlisting}
	\caption{Pseudo code for the first stage of the SG  experiment and the concept of an oven, in Chapter 1 of Nielsen's \cite{nielsen_chuang_2010} quantum information processing (Figure 1.22 and equations 1.56 through 1.59) where a state ie passed through the Pauli Z gate.}
	\label{fig:sg:0}
\end{figure}

Using the SG apparatus oriented in a particular direction, we can prepare electrons to have spins in that direction \cite{Zhu2011}. We know that performing a measurement changes the state of a quantum system, here we can assume that the state was created by passing a particle through an S-G ($z$ axis) apparatus oriented in the Z direction.

 If we measure $S_{z}$ (choosing the $Z$ axis for our measurements is common practice) on a particle in the general state $Z$ , then the possible outcomes are $z_{+}$ 1, with probability $|a|^{2}$, or $z_{-}$ 0 ($-\hbar/2$), with probability $|b|^{2}$ \cite{Zhu2011}.:
 \begin{equation}
 	|a|^{2}+|b|^{2}=1
 \end{equation}

 In this case the state of the spin $\frac{1}{2}$ particle equals $z_{+}$ or $s$ and therefore the bit "0":
 $$
 z_{+}=\left(\begin{array}{l}
 	1 \\0 \end{array}\right)
 $$
will always be observed with absolute certainty. Therefore $\alpha=1$, $\beta=0$ and the wave-function in the $z_{+}$, $z_{-}$ basis is:

\begin{equation}
	|\psi\rangle=\alpha|0\rangle+\beta|1\rangle=1*|z_{+}\rangle +0*|z_{-}\rangle=|z_{+}\rangle
\end{equation}

In \cite{Zhu2011} \textit{Improving students’ understanding of QM via the Stern–Gerlach experiment}  the authors draw unambiguous conclusions that align with our suggestions.:\\
\textit{``The preparation of a specific quantum state may be challenging to achieve in the laboratory but it is relatively easy to conceptualize theoretically at least in a 2D Hilbert space with SGE. We find that the students have difficulty with the preparation of a specific quantum state even in a 2D Hilbert space.'' }
\\
and:\\
\textit{``Here, we discuss investigation of students' difficulties about the SG experiment by giving written tests and interviewing advanced undergraduate and graduate students in QM courses. We also discuss preliminary data from two QM courses that suggest that a Quantum Interactive Learning Tutorial (QuILT) related to the SG experiment is helpful in improving students' understanding of these concepts.''}

\subsubsection{The crucial takeaway for CS students}
Finally, for the student, it is important to understand the following concept that \textit{\textbf{relates an operator to it's eigenvalues}}. The 4th Postulate of QM states:
\textbf{The probability of obtaining the eigenvalue $a_{\mathrm{n}}$ in a measurement of the observable $A$ on the system in the state $|\psi\rangle$ is
	$$
	\mathcal{P}\left(a_{n}\right)=\left|\left\langle a_{n} \mid \psi\right\rangle\right|^{2},
	$$
	where $\left|a_{n}\right\rangle$ is the eigenvector of $A$ corresponding to the eigenvalue $a_{\mathrm{n}}$.} \\

\begin{figure}[H]
	\centering
	\includegraphics[totalheight=2cm]{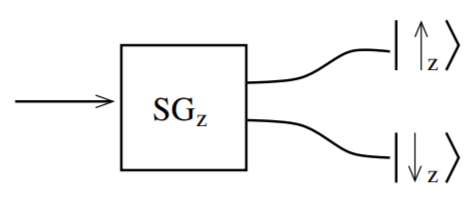}
	\caption{The figure depicts the SG experiment oriented in the z-direction (SGz). Source \cite{babis}.}
	\label{fig::sg02}
\end{figure}

Here we have an arbitrary linear combination of the basis states, where $\alpha, \beta \in \mathbb{C}$.
The state of the electron spin after exiting the oven (See Fig. \ref{fig::sg02}) is:
\begin{equation}
	|\psi\rangle=\alpha|z_{+}\rangle+\beta|z_{-}\rangle
\end{equation}

Or equivalently:
\begin{equation}
	|\psi\rangle=\longleftrightarrow \alpha\left(\begin{array}{l}
		1 \\
		0
	\end{array}\right)+\beta\left(\begin{array}{l}
		0 \\
		1
	\end{array}\right)=\left(\begin{array}{l}
		\alpha \\
		\beta
	\end{array}\right)
\end{equation}

And also $\langle\psi \mid \psi\rangle=|\alpha|^{2}+|\beta|^{2}=1$.

However, $\left|\alpha_{\uparrow}\right|^{2}=\left|\beta_{\downarrow}\right|^{2}$ does \emph{not} necessarily hold. Therefore, when exiting the first S-G ($z$ axis) (Fig.\ref{fig::sg01}), the electron collapses to $|\uparrow\rangle$ or $|\downarrow\rangle$, with probabilities $\left|\alpha_{\uparrow}\right|^{2}$ and $\left|\beta_{\downarrow}\right|^{2}$, respectively. Mathematically:
\index{Probability amplitude}
We can first compute the inner product of this ket $|\psi\rangle$ with the bra $|z_{+}\rangle$:
\begin{equation}
	\begin{aligned}
		\langle z_{+} \mid \psi\rangle &=\langle z_{+} |\alpha|z_{+} \rangle+\langle z_{+} |\beta|z_{-} \rangle \\
		&=\alpha\langle z_{+} \mid z_{+} \rangle+\beta\langle z_{+} \mid z_{-} \rangle =\alpha
	\end{aligned}
\end{equation}

Then, the probability that a measurement yields a certain result is the square of the modulus of the corresponding amplitude \cite{shoshani2021}:

\begin{equation}
	\begin{aligned}
		&\mathcal{P}(z_{+})=\left|\left\langle z_{+} \mid \psi\right\rangle\right|^{2}=|\alpha|^{2}
	\end{aligned}
\end{equation}

\emph{This is the probability that the state $|\psi\rangle$ is found to be in the state $|z_{+}\rangle$ when a measurement of $S_{z}$ is made.} Likewise:
\begin{equation}
	\mathcal{P}(z_{-})=|\beta|^{2}.
\end{equation}
Here, we used the property that scalars can be moved freely through either bras or kets.
\index{Bra}
\index{Ket}
\subsection{Using Julia for teaching quantum simulation}

We believe that Julia's symbolic variable ($\Delta * \sigma_\mathrm{x}$) and operator ($\otimes$) support~\cite{Gawron2018,kramer2018quantumoptics} make it an ideal choice for turning a theoretical quantum mechanic concept or formula into code \cite{julia2015}. For scientific computing, the high-level just-in-time compiled language Julia is extremely effective and while there were numerous attempts at writing Julia-based QM and quantum information processing libraries, only a few survived, some of which we present here:.
\begin{itemize}[leftmargin=5.5mm]
\item QuantumInformation.jl \cite{Gawron2018} for instance, whose development efforts ceased a few years ago, is among these. It makes heavy usage of symbolic representations as exemplified in Fig. (\ref{fig:qi:0}):
\begin{figure}[H]
	\begin{lstlisting}[caption={},language=Ada,]
	$\phi=1 / \operatorname{sqrt}(2) *(\operatorname{ket}(1,4)+\operatorname{ket}(4,4))$
	$\xi=\operatorname{ptrace}(\operatorname{proj}(\phi),[2,2],[2]$,
	@test $\xi \approx I / 2$ atol=1e-15
	\end{lstlisting}
\caption{Julia \href{https://github.com/iitis/QuantumInformation.jl/blob/75899f9de3453ad22191199595d75ade5a12dddc/test/ptrace.jl}{code snippet}  depicting the \textbf{usage of Unicode }(e.g. $\phi$, $\xi$ etc.) symbols in place of variable names. In this case, the application of partial trace to two Kets.}
\label{fig:qi:0}
\end{figure}

\item QuantumOptics.jl \cite{kramer2018quantumoptics} is another Julia framework for simulating open quantum systems, although judging from a cursory perusal of the discussion groups, the primary audience is physicists and not computer scientists. It also utilises \href{https://github.com/qojulia/QuantumOptics.jl-examples}{symbolic representations} even more heavily
 as exemplified in Fig. (\ref{fig:qi:1}):
\begin{figure}[H]
	\begin{lstlisting}[caption={},language=Ada,]
		using QuantumOptics
		$\Omega =0.5$
		$t =[0: 0.1: 10 ;]$
		$b=$ SpinBasis $(1 / / 2)$
		$H=\Omega *(\operatorname{sigmap}(b) \otimes \operatorname{sigmam}(b)+\operatorname{sigmam}(b) \otimes \operatorname{sigmap}(b))$
	\end{lstlisting}
	\caption{Line number $6$ in the code snippet, \textbf{is a marvellous example of the expressive power of Julia}, and how a mathematical concept is translated almost one-to-one to an actual code realization.}
	\label{fig:qi:1}
\end{figure}

\index{Yao.jl}
\index{Julia}

\item Yao.jl \cite{Luo2020Yao} is a very active and highly maintained quantum computing library in Julia which is described by it's authors as \textit{an open source framework that aims to empower quantum information research with software tools } \cite{Luo2020Yao}.  In a working Julia environment, the following code snippet Fig. (\ref{fig:jul:003}) can be used to install the dependencies required for running Yao.jl
\begin{figure}[H]
	\begin{lstlisting}[caption={},language=C++,]
		begin
			using Pkg
			Pkg.activate(mktempdir())
			Pkg.Registry.update()
			Pkg.add("Yao")
			Pkg.add("YaoPlots")
			Pkg.add("StatsBase")
			Pkg.add("Plots")
			Pkg.add("BitBasis")
		end
	\end{lstlisting}
	\caption{Installation instructions for Yao.jl}
	\label{fig:jul:003}
\end{figure}
\index{Yao.jl}
\index{Julia}
QuantumBFS, an open source organization for quantum science, is responsible for the project. Xiu-Zhe (Roger) Luo and Jin-Guo Liu, who are members of the organization, have been actively answering our questions on the discussion boards. For instance, upon our request they added a feature to the library that allows \href{https://github.com/QuantumBFS/YaoPlots.jl/pull/60}{plotting with barriers}. A simple quantum circuit created using Yao.jl is depicted in Fig. \ref{fig:jul:004}:
\index{Quantum circuit}
\begin{figure}[H]
	(a)
	\begin{lstlisting}[caption={},language=C++,]
		cr = chain(2,
		kron(H, X),
		kron(1=>Y, 2=>H),
		);
	plot(cr)
	\end{lstlisting}
(b) $$\adjustbox{scale=1}{%
	\begin{tikzcd}
		\lstick{$q_0$} & \circuitH & \circuitX & \qw \\
		\lstick{$q_1$} & \circuitH & \circuitH & \qw \\
	\end{tikzcd}
}$$
(c) $$ (H \otimes X) \times (X \otimes X) $$
	\caption{(a) Creating a simple quantum circuit in Yao.jl. (b) the respective circuit diagram in YaoPlots.jl (c) the actual mathematical representation of the tensor product.}
	\label{fig:jul:004}
\end{figure}
\index{Quantum circuit}
For the purposes of academic research and the incentives mentioned, we decided to conduct our experiments in Julia using a variety of libraries, including Yao.jl and PastaQ.jl. To help with visualization and plotting, we also used quirk \cite{quirk2016}, a tool that allows users to easily create quantum circuit simulations, and YaoPlots.jl \cite{Wang2020}, a plotting extension specifically designed for visualizing circuits created in Yao.jl.

\index{Yao.jl}
\index{Julia}

\end{itemize}

%

\section{The classical simulation of Entangled-states }
\index{Yao.jl}
\index{Julia}
\index{Entanglement}
After going over some of the theoretical underpinnings of entanglement in Sec. (\ref{ent-th}), we then proceed to utilise Yao.jl to simulate entanglement in the following quantum information processing \cite{nielsen_chuang_2010} experiments:
\begin{enumerate}
\item \raisebox{-0.2em}{\textcolor{cavitycolor}{\rule{0.8em}{0.8em}}} Superposition via the Hadamard (\ref{sub:sup:had})
\index{Hadamard}
\item \raisebox{-0.2em}{\textcolor{cavitycolor}{\rule{0.8em}{0.8em}}} Bell-state generation (\ref{sub:sup:had1})
\item \raisebox{-0.2em}{\textcolor{cavitycolor}{\rule{0.8em}{0.8em}}} GHZ state generation (\ref{sub:sup:had2})
\item \raisebox{-0.2em}{\textcolor{cavitycolor}{\rule{0.8em}{0.8em}}} Hardy's state generation (\ref{sub:sup:had3})
\end{enumerate}

\subsection{The theory of Entanglement}
\index{Entanglement}
\label{ent-th}
Given the extensive literature on entanglement \cite{woo,shoshani2021,nielsen_chuang_2010, Levesque2007}, we will not delve into its theoretical aspects in this context. Instead, our focus will be on teaching entanglement through quantum programming. Entanglement is a phenomenon in which two qubits are correlated, and it has been very helpful in the field of quantum computing, such as in Shor's and Jozsa's algorithms ~\citep{jozsa2003}. When two qubits are entangled, measuring one of them causes the other to take on a specific state. This means that the measurement of one qubit can influence the potential behaviour of the second qubit. In other words, the statistical connections between the measurements of entangled quantum states cannot be explained by local realist physical theories. It is often beneficial for students to understand how to test for entanglement ~\citep{huang2019}, as we will discuss in a later section \ref{Von-neumann}. when using the Von Neumann entropy measure.

\index{Von-Neumann}
\index{Von-Neumann entropy}
\index{Entangelment qunatification}

Let us now shortly introduce the meaning of entanglement. A state of a bipartite system is referred to as entangled if it cannot be expressed as the direct product of two states from the two subsystem Hilbert spaces \cite{nielsen_chuang_2010}. Mathematically this implies that:
\begin{equation}
	\Psi_{A \otimes B} \neq \Psi_A \otimes \Psi_B
\end{equation}

Therefore, in principle, one could:
\begin{itemize}[leftmargin=5.5mm]
\item Generate a system of two physically local, entangled qubits, such as any form of the Bell pairs.
\item Separate them by a significant distance.
\item Measure one of the qubits, resulting in the collapse of the entire entangled system (i.e., both qubits).
\item Immediately thereafter, measure the second qubit and provide a result that is correlated with the first qubit.
\end{itemize}

\subsection{Entanglement generation using Julia}
\index{Entanglement}
\index{Superposition}
\index{Pauli X}
\index{Pauli Z}
Quantum programmes make considerable use of superposition and entanglement, which are the fundamental basis for quantum computing's computational advantage over traditional computing \cite{nielsen00,woo}.  A quantum programme is a series of quantum operations (gates) executed on a group of qubits and single-qubit gates such as the Hadamard. In order to take advantage of quantum parallelism, Hadamard gates are commonly used in quantum information processing to set the input qubits in the \textbf{uniform superposition} state, $|+\rangle=\frac{1}{\sqrt{2}}|0\rangle+\frac{1}{\sqrt{2}}|1\rangle$ which  switches between the Pauli $X$ and $Z$.
\index{Hadamard}

\subsubsection{The Hadamard and superposition generation}
\label{sub:sup:had}
\index{Hadamard}
The superposition principle serves a key role in the theory of quantum information processing, and one of the most famous experiments in quantum superposition, is the double-slit experiment \cite{shoshani2021} consisting of a source, a double-slit assembly,and an observation screen used to observe interference fringes. A \textbf{single} Hadamard transformation plays the role of a  $50/50$ beamsplitter while a Mach-Zehnder interferometer  consists of  a sequence of \textbf{two} Hadamard transformations \cite{shoshani2021}.
\index{Double-slit}
\index{Beamsplitter}
\index{Mach-Zehnder interferometer }

\begin{figure}[H]
		\centering
	\begin{equation}
	(a)	H=\frac{1}{\sqrt{2}}[(|0>+| 1>)<0|+(|0>-| 1>)<1|]
	\label{eq:hadamard0}
	\end{equation}
	\begin{equation}
   (b) H=\frac{1}{\sqrt{2}}\left[\begin{array}{rr}
			1 & 1 \\
			1 & -1
		\end{array}\right]
		\label{eq:hadamard1}
	\end{equation}
(c)
\begin{figure}[H]
	\centering
	\includegraphics[totalheight=5cm]{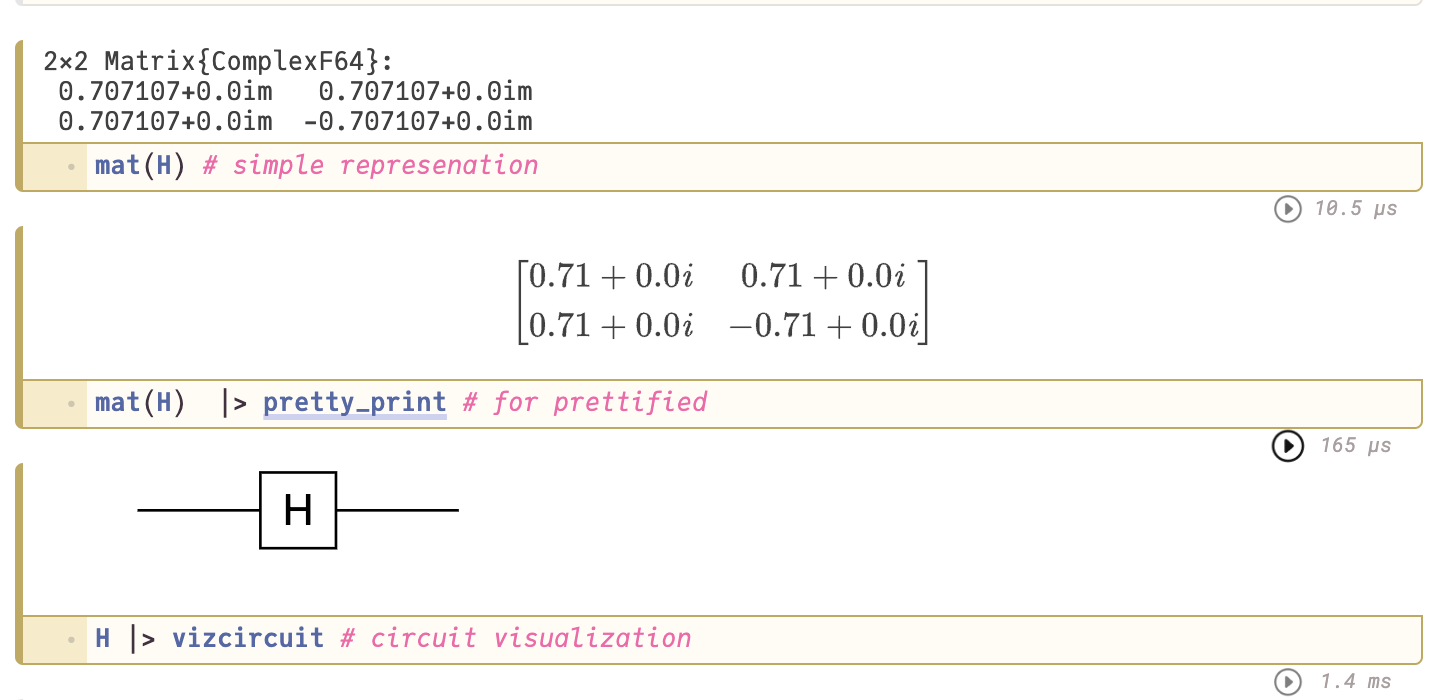}
\end{figure}
	\caption{The 1-qubit Hadamard. (a) The Hadamard in the Dirac notation (b) The Hadamard  in a matrix notation. (c) using Julia to generate a Hadamard gate.}
	\label{fig:hadamard:gate}
\end{figure}

\textbf{Physicists} are thought that a physical process is related to a general unitary matrix via the unitary evolution equation $U=\exp (-i \mathcal{H} t / \hbar)$ which is expressed as\adjustbox{scale=0.45}{
\begin{tikzcd}\lstick{} & \circuitU & \qw \\\end{tikzcd}} in a quantum circuit diagram. However, as a first step in the study of entanglement and Bell-state preparation \cite{woo}, the Hadamard may be introduced to\textbf{ computer science }students via it's two forms as depicted in Fig. (\ref{fig:hadamard:gate}) and expressed as follows in both the Dirac Eq.\eqref{eq:hadamard0} and matrix Eq.\eqref{eq:hadamard1} notations:
\index{Quantum circuit}
\index{Hadamard}
\index{Unitary}
\index{Unitary time evolution}
\index{Dirac}
The result of applying the Hadamard gate \cite{nielsen00} \adjustbox{scale=0.7}{%
	\begin{tikzcd}
		\lstick{$\Psi$} & \circuitH & \qw \\
	\end{tikzcd}
} to the quantum states $|0\rangle$ and $|1\rangle$ is depicted in Fig. (\ref{fig:had:out2}):
\begin{figure}[H]
	\index{Hadamard}

	(a) $$\adjustbox{scale=1}{%
		 \begin{quantikz}[slice all,slice style={shorten <=-0.1cm,
				shorten >=-0.1cm},slice label style={yshift=0.1cm}]
			\lstick{$\ket{x}$} 	&\gate[style={fill=cyan!20}]{H}  & \rstick{$(-1)^x|x\rangle+|1-x\rangle, |x\rangle=\{|0\rangle,|1\rangle\}$}\qw
		\end{quantikz}
	}$$
	(b) \begin{equation}
		\begin{aligned}
			&H|0\rangle=\frac{1}{\sqrt{2}}(|0\rangle+|1\rangle) \equiv|+\rangle_x \\
			&H|1\rangle=\frac{1}{\sqrt{2}}(|0\rangle-|1\rangle) \equiv|-\rangle_x
		\end{aligned}
		\label{fig:hadamard22:gate}
	\end{equation}
(c)
	\begin{lstlisting}[caption={},language=Ada]
	begin
			st0 = normalize!(arrayreg(bit"0"))
			state(st0) |> pretty_print
			r0 = apply!((st0), H)
			state(r0)
	end
	> julia Matrix{ComplexF64}:
		0.7071067811865475 + 0.0im
		0.7071067811865475 + 0.0im
\end{lstlisting}
	\caption{(a) The effect of the gate $H$ acting on a qubit in state $|x\rangle$.  (b) The resulting quantum states (c) Yao.jl code snippet depicting the application fo the H gate on the two basic states in the computational basis. Applying the Hadamard to quantum state $|0\rangle$ for instance, results in $H|0\rangle=\frac{1}{\sqrt{2}}(|0\rangle+|1\rangle)$.}
	\label{fig:had:out2}
	\index{Hadamard}
\end{figure}
\index{Yao.jl}
\index{Julia}
\index{Basis}
Using Yao.jl, the full processes for generating the two quantum states in superposition is depicted in Fig. (\ref{fig:hadamard2:gate}):
\index{Superposition}
\begin{figure}[H]
	\centering
	\begin{figure}[H]
		\centering
		\includegraphics[totalheight=8cm]{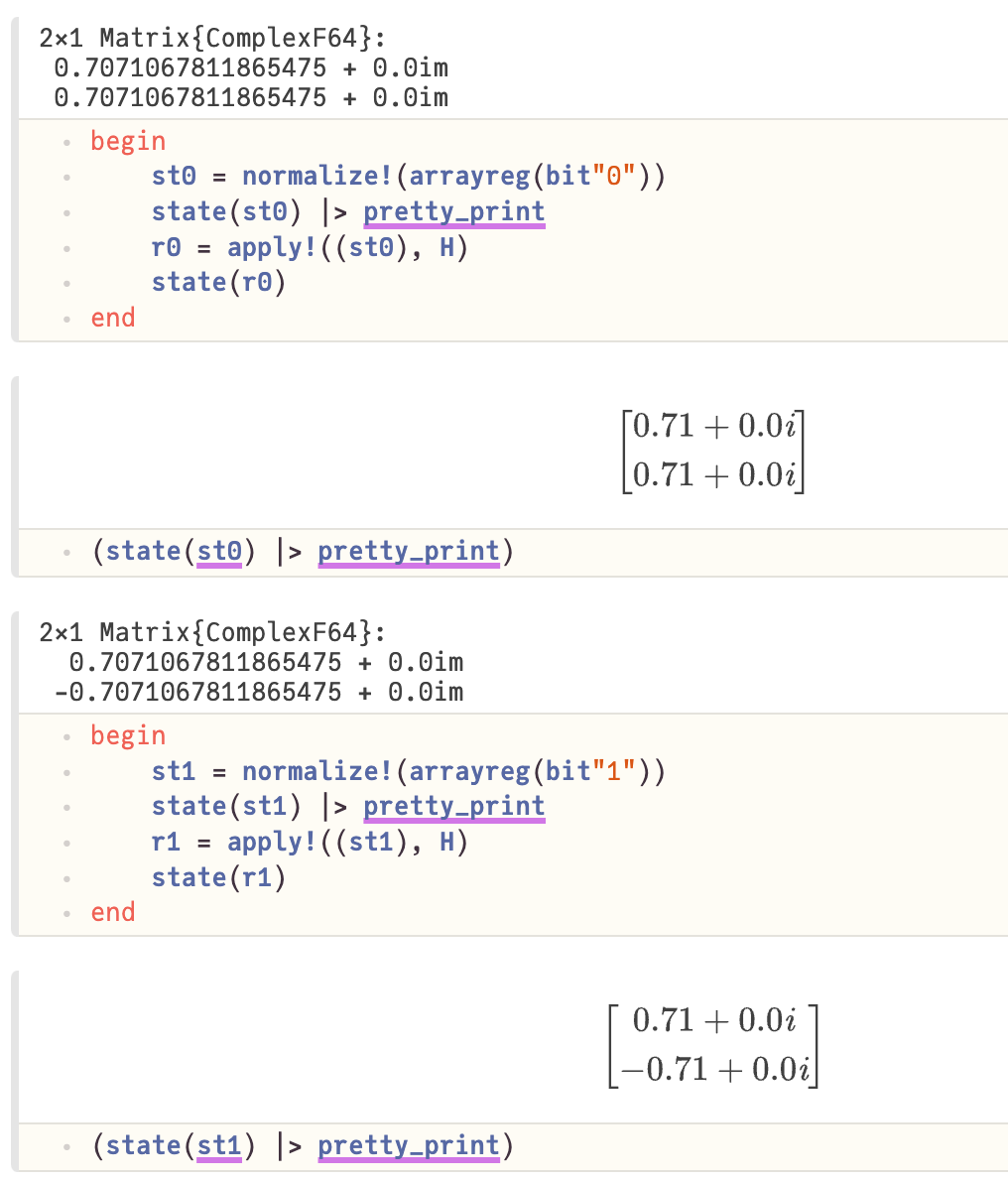}
	\end{figure}
	\caption{Julia code snippet for generating superposition via the Hadamard, matching the values in the mathematical expression in Eq.\eqref{fig:hadamard22:gate}.}
	\label{fig:hadamard2:gate}
\end{figure}

It is important for the students to understand that while \textbf{\textit{the Hadamard gate can create quantum superposition in a single qubit}} (such as $H|0\rangle=|+\rangle$), it cannot be used to entangle multiple qubits. No single-qubit gate has this capability.
\index{Hadamard}

\subsubsection{Using the Hadamard for Entanglement generation}
\label{sub:had:ent}
\index{Superposition}
We have already stated that the H gate alone can not generate an entangled state. The entangling operation consists of a local Hadamard transformation which maps $|0\rangle \rightarrow(|0\rangle+|1\rangle) / \sqrt{2},|1\rangle \rightarrow(|0\rangle-|1\rangle) / \sqrt{2}$ followed by the application of a local CNOT (Controlled NOT).  We remind the reader that a two-qubit CNOT operator acts as follows:
\begin{equation}
	\text { CNOT }|00\rangle=|00\rangle, \quad \text { CNOT }|01\rangle=|01\rangle, \quad \text { CNOT }|10\rangle=|11\rangle, \quad \text { CNOT }|11\rangle=|10\rangle .
\end{equation}
Where the 0-th qubit is the control qubit and the 1-st qubit is the target qubit.
\begin{figure}[H]
	(a)
	\begin{lstlisting}[caption={},language=Ada]
		begin
		st4 = normalize!(arrayreg(bit"00"))
		state(st4) |> pretty_print
		cir=chain(2,
		kron(H,I2),
		control(1,2=>X))
		r4 = apply!((st4), cir)
		state(r4) |> pretty_print
		end
		>julia
		$\left[\begin{array}{c}0.71+0.0 i \\ 0.0 i \\ 0.0 i \\ 0.71+0.0 i\end{array}\right]$
	\end{lstlisting}
	(b)
	\begin{quantikz}[slice all,remove end slices=1,slice
		titles=$\Psi$ \col,slice style=blue,slice label style
		={inner sep=1pt,anchor=south west,rotate=30}]
		& \gate{H} & \ctrl{1} & \meter{} && \\
		& \qw & \targ{}  & \meter{} &&
	\end{quantikz}
	\caption{(a) Yao.jl code snippet depicting the application fo the H gate on the two basic states in the computational basis. (b) In a step by step approach, following the application of H gate (e.g. \textbf{$\Psi 1$ }), the quantum state turns into $|+\rangle \otimes|0\rangle=\frac{|0\rangle+|1\rangle}{\sqrt{2}} \otimes|0\rangle=\frac{(|0\rangle+|1\rangle) \otimes|0\rangle}{\sqrt{2}}=\frac{|0\rangle \otimes|0\rangle+|1\rangle \otimes|0\rangle}{\sqrt{2}}=\frac{|00\rangle+|10\rangle}{\sqrt{2}}$.} Subsequently in \textbf{$\Psi 2$},  the CNOT gate  maps the state $\frac{|00\rangle+|10\rangle}{\sqrt{2}}$ to the state $\frac{|00\rangle+|11\rangle}{\sqrt{2}}$. The full mathematical derivation is depicted in Eq.\eqref{had:cnot}.
	\label{fig:had:out3}
	\index{Yao.jl}
	\index{Julia}
\end{figure}
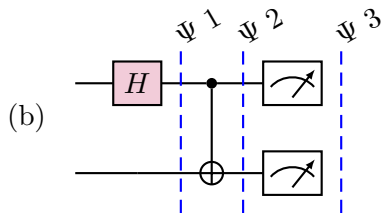
\textbf{\raisebox{-0.2em}{\textcolor{orangebdark}{\rule{0.8em}{0.8em}}} Paper-and-pencil computation:}\\
In matrix form the CNOT is represented as:
\begin{equation}
	{CNOT} \rightarrow\left(\begin{array}{cccc}
		1 & 0 & 0 & 0 \\
		0 & 0 & 0 & 1 \\
		0 & 0 & 1 & 0 \\
		0 & 1 & 0 & 0
	\end{array}\right)
\end{equation}
\index{Entanglement}
\index{CNOT}
Here in Fig. (\ref{fig:had:out3}) we apply an entangling operation, an Hadamard gate followed by a CNOT gate where Initially, the 2-qubit state of the system is $|0\rangle \otimes|0\rangle=|00\rangle$.:
\index{Hadamard}
\begin{equation}
	|00\rangle \rightarrow|0\rangle \otimes H|0\rangle=\frac{|00\rangle+|01\rangle}{\sqrt{2}} \rightarrow C N O T_{01} \frac{|00\rangle+|01\rangle}{\sqrt{2}}=
\end{equation}
\begin{equation}
	\frac{C N O T_{01}|00\rangle+C N O T_{01}|01\rangle}{\sqrt{2}}=\frac{|00\rangle+|11\rangle}{\sqrt{2}}
\end{equation}
\index{CNOT}
\index{Controll}
\index{Target}
\begin{equation}
	\begin{aligned}
		\Psi 2= \operatorname{CNOT}\left(\frac{|00\rangle+|10\rangle}{\sqrt{2}}\right) & =\left(\begin{array}{llll}
			1 & 0 & 0 & 0 \\
			0 & 1 & 0 & 0 \\
			0 & 0 & 0 & 1 \\
			0 & 0 & 1 & 0
		\end{array}\right)\left(\frac{|00\rangle+|10\rangle}{\sqrt{2}}\right) \\
		& =\frac{1}{\sqrt{2}}\left(\begin{array}{llll}
			1 & 0 & 0 & 0 \\
			0 & 1 & 0 & 0 \\
			0 & 0 & 0 & 1 \\
			0 & 0 & 1 & 0
		\end{array}\right)(|00\rangle+|10\rangle) \\
		& =\frac{1}{\sqrt{2}}\left(\begin{array}{llll}
			1 & 0 & 0 & 0 \\
			0 & 1 & 0 & 0 \\
			0 & 0 & 0 & 1 \\
			0 & 0 & 1 & 0
		\end{array}\right)\left(\left(\begin{array}{l}
			1 \\
			0 \\
			0 \\
			0
		\end{array}\right)+\left(\begin{array}{l}
			0 \\
			0 \\
			1 \\
			0
		\end{array}\right)\right) \\
		& =\frac{1}{\sqrt{2}}\left(\begin{array}{llll}
			1 & 0 & 0 & 0 \\
			0 & 1 & 0 & 0 \\
			0 & 0 & 0 & 1 \\
			0 & 0 & 1 & 0
		\end{array}\right)\left(\begin{array}{l}
			1 \\
			0 \\
			1 \\
			0
		\end{array}\right)
	\end{aligned}
\label{had:cnot}
\end{equation}

Which equals:
\begin{equation}
	=\frac{1}{\sqrt{2}}\left(\left(\begin{array}{l}
		1 \\
		0 \\
		0 \\
		0
	\end{array}\right)+\left(\begin{array}{l}
		0 \\
		0 \\
		0 \\
		1
	\end{array}\right)\right)
=\frac{1}{\sqrt{2}}(|00\rangle+|11\rangle)
\label{had:cnot:final}
\end{equation}
This also equals the value of the matrix generated by Yao.jl in line $11$ in Fig. \ref{fig:had:out2} (b).
\index{Yao.jl}
\index{Julia}
\index{Quantum simulation}
\index{Shots}
We now simulate this simple circuit on a classical computer by running 1024 shots:
\index{Quantum circuit}
\begin{figure}[H]
	(a)
	\begin{lstlisting}[caption={},language=Ada]
		measuredqubits = r4 |> r->measure(r, nshots=1024)
		plotmeasure(measuredqubits)
	\end{lstlisting}
	(b)
	\begin{figure}[H]
		\centering
		\includegraphics[totalheight=5cm]{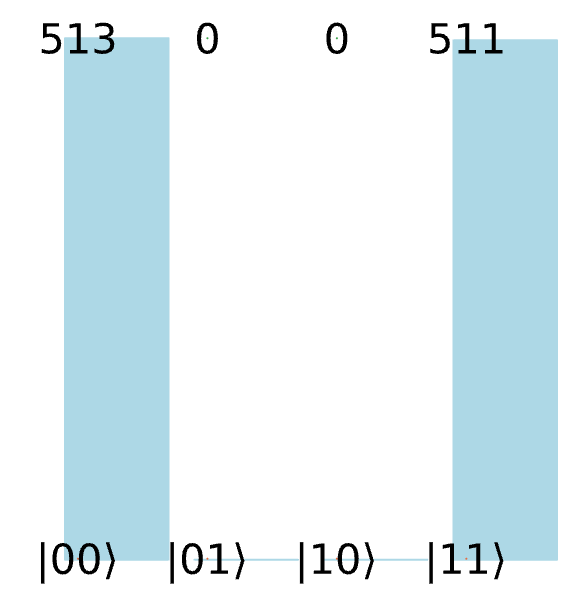}
	\end{figure}
	\caption{(a) Yao.jl code snippet for running a simulation of 1024 shots on a classical computer. (b) visualizing the results using a Yao.jl bar plot. This finding is not unexpected given that Yao's classical simulator reproduces the quantum equations perfectly. In particular, the initial measurement (which collapses the entanglement) will, on average, choose between $|00\rangle$ and $|11\rangle$ for the collapsed state. On actual quantum computing hardware, which is susceptible to quantum mistakes as a result of decoherence and noise, the results may be far from ideal.}
	\label{fig:had:sim}
	\index{Yao.jl}
	\index{Julia}
\end{figure}

How can we confirm that a state is \textbf{\textit{truly entangled}}? One method is to use the von-Neumann entropy measure \ref{Von-neumann}, which is a commonly used measure of entanglement.
\index{Entanglement}
\index{von-Neumann entropy}
\index{Entanglement entropy}
\index{Entropy}
\index{Pure states}
\index{Mixed states}
\subsection{The von-Neumann entropy measure}
\label{Von-neumann}

In this section, density matrices are frequently utilized and therefore, we  provide a brief overview of their characteristics. For a more detailed treatment of the topic, interested readers can refer to external sources, such as \cite{shoshani2021,nielsen_chuang_2010}.

\subsubsection{Density matrices}
\index{Density matrices}
Density matrices are a tool that can be used to describe quantum states in a concise and effective manner, particularly when dealing with statistical mixtures of different states. They are commonly used in the study of quantum systems.

In the context of entropy, the density matrix of a pure state $\{(1,|\psi\rangle)\}$ is given by $\rho=|\psi\rangle\langle\psi|$ while for a mixed state $\rho=\sum_i p_i\left|\psi_i\right\rangle\left\langle\psi_i\right|$. By definition, the density matrix is a positive, Hermitian matrix with a trace of one.
\index{Purity}
\index{Hermitian}
\index{Density matrices}
We remind the reader that a\textit{ pure state} is a quantum state whose wave-function is known with absolute certainty while a \textit{mixed state} is a statistical distribution of pure states \cite{shoshani2021} and hence pure states have purity one and mixed states have purity of less than one.\cite{nielsen_chuang_2010}.

Here the notion of purity also arises,  and is described by the density operator $\rho$ which is defined to be $\operatorname{Tr}\left[\rho^2\right]$. The von Neumann entropy is defined as:
\begin{equation}
	S(\rho)=-\sum_i^m \lambda_i \log \lambda_i=\operatorname{tr}(-\rho \log \rho)
\end{equation}

For a pure quantum state the following condition holds, $S(\rho)=0$, for a maximally mixed state $\rho=\frac{\mathbf{I}}{m}$ and $S(\rho)=\log m$. The state $\rho_{A B}$ is entangled if $S\left(\rho_A \mid \rho_B\right)<0$.

\index{Entropy}
\index{Maximally mixed states}
\index{Paper-and-pencil}
\index{Density matrices}
\textbf{\raisebox{-0.2em}{\textcolor{orangebdark}{\rule{0.8em}{0.8em}}}  Paper-and-pencil computation:}\\
Now assume a system is in the pure state $|\psi\rangle=\frac{1}{\sqrt{2}}(|00\rangle+|11\rangle)$ (Eq. \eqref{had:cnot:final}). The reduced density matrix state is mixed, even though the state $|\Psi\rangle$ itself is pure (Eq. \eqref{desnsityA}):
\begin{equation}
	\begin{aligned}
		\rho_{A B} & =|\Psi\rangle\langle\Psi|=\frac{1}{2}(|00\rangle\langle 00|+| 00\rangle\langle 11|+| 11\rangle\langle 00|+| 11\rangle\langle 11|) \\
		\operatorname{tr}_B\left(\rho_{A B}\right) & =\frac{1}{2}(|0\rangle\langle 0|+| 1\rangle\langle 1|)=\frac{\mathbb{1}_A}{2} .
	\end{aligned}
\label{desnsityA}
\end{equation}
and therefore it's density matrix is:
\index{Density matrix}
\index{Bell state}
\begin{equation}
	\rho=|\psi\rangle\langle\psi|
\end{equation}
For our case:
\begin{equation}
	\rho=\frac{1}{2}\left(\begin{array}{llll}
		1 & 0 & 0 & 1 \\
		0 & 0 & 0 & 0 \\
		0 & 0 & 0 & 0 \\
		1 & 0 & 0 & 1
	\end{array}\right)
\end{equation}

Or in Yao.jl (\ref{fig:density:yao}):
\index{Yao}
\begin{figure}[H]
(a)
\begin{lstlisting}[caption={},language=Ada]
	begin
			reg = (arrayreg(bit"00") + arrayreg(bit"11")) / sqrt(2)
			(statevec(reg) * statevec(reg)') |> pretty_print
	end
	> julia
	$\left[\begin{array}{cccc}0.5+0.0 i & 0.0 i & 0.0 i & 0.5+0.0 i \\ 0.0 i & 0.0 i & 0.0 i & 0.0 i \\ 0.0 i & 0.0 i & 0.0 i & 0.0 i \\ 0.5+0.0 i & 0.0 i & 0.0 i & 0.5+0.0 i\end{array}\right]$
\end{lstlisting}
(b)
\begin{lstlisting}[caption={},language=Ada]
	begin
			rho = density_matrix(reg)
			(von_neumann_entropy(rho))
	end
> julia
2.4231940935448884e-14 (e.g.=0)
\end{lstlisting}
	\caption{(a) Calculating the density matrix for the pure quantum bell-state $|\psi\rangle=(|00\rangle+|11\rangle) / \sqrt{2}$ using Yao.jl. (b) The entropy of a quantum system is computed by deriving the eigenvalues $\lambda_n$ of the density operator $\rho$ and preforming the operation $S=-\sum_n \lambda_n \log \left(\lambda_n\right)$. By expressing the operator as a matrix and calculating its eigenvalues, the Von Neumann entropy is realised in Yao.jl where subsequently, the eigenvalues are plugged into the equation for Von Neumann entropy. This is demonstrated by the  second code fragment.}
	\label{fig:density:yao}
\end{figure}
\index{Density matrix}

It is a well-established fact that a pure state in a quantum system has no entropy, as the only non-zero eigenvalue of the density matrix in a pure state is $\lambda=1$. When this eigenvalue is applied to the Von Neumann entropy equation, it becomes clear that a pure state, being a completely known system, should not possess any entropy. This is in line with the general understanding of entropy and pure states in the quantum world. This idea has been widely accepted in the field, as evident from the various sources cited in \cite{shoshani2021, woo, nielsen_chuang_2010}.
\index{Eigenvalue}

\subsection{Bell-state generation}
\label{sub:sup:had1}
\index{Bell states}
\index{Superpositions}

The Bell states are superpositions of two particles that are maximally entangled, as described in a study by \cite{shoshani2021}. These states occur when two spin-1/2 particles are produced at the same time and can be represented by four wave functions that encompass the entire spin states of the particles. The overall system must be represented by a four-vector due to the two possible spin orientations for each particle.

In section (\ref{sub:had:ent}) we have already studied the generation of the most well-known Bell-state $|\Phi^{+}\rangle$:
\begin{equation}
	|\Phi^{+}\rangle=\frac{1}{\sqrt{2}}(|00\rangle+|11\rangle)=\left[\begin{array}{c}
		\frac{1}{\sqrt{2}} \\
		0 \\
		0 \\
		\frac{1}{\sqrt{2}}
	\end{array}\right]
\end{equation}
\index{Quantum circuit}
Using a quantum circuit comprised of a Hadamard (H) gate, an identity (I) gate, and a controlled-not gate (CNOT) as described in \ref{sub:had:ent}, any Bell state may be created from two classical bits:
\begin{figure}[H]
	(a)
	\begin{quantikz}[slice all,remove end slices=1,slice
		titles=$\Psi$ \col,slice style=blue,slice label style
		={inner sep=1pt,anchor=south west,rotate=30}]
		&   \gate{H}  & \ctrl{1}  & \meter{} && \\
		&  \qw & \targ{0}  & \meter{} &&
	\end{quantikz}

	(b) \begin{equation}
		\operatorname{CNOT}(H \otimes I)|01\rangle=\frac{1}{\sqrt{2}} \cdot\left(\begin{array}{l}
			1 \\
			0 \\
			0 \\
			-1
		\end{array}\right)
	\end{equation}
(c) \begin{lstlisting}[caption={},language=Ada]
	cr=chain(2, kron(H,I2), control(1,2=>X))
	$\psi 0=$ ArrayReg(bit"01") $\mid>$ normalize!$
	$\psi 0=(\psi 0 \mid>$ cr $)$
	$@show ((\psi 0$.state))$
\end{lstlisting}
	\caption{(a) The Bell-state generation circuit (b) The mathematical expression for generating a Bell-state (b) Yao.jl code snippet for creating the circuit and measuring the state.}
	\label{fig:output_matrix}
\end{figure}
\index{Quantum circuit}
The result of running the circuit in \ref{fig:output_matrix} on a classical computer is presented in \eqref{eq:yao:bell-gen}:
\begin{equation}
	\begin{array}{r}
		4 \times 1 \text { Matrix }\{\text { ComplexF } 64\} \\
		0.7071067811865475+0.0 \mathrm{im} \\
		0.0+0.0 \mathrm{im} \\
		0.0+0.0 \mathrm{im} \\
		-0.7071067811865475+0.0 \mathrm{im}
	\end{array}
\label{eq:yao:bell-gen}
\end{equation}

For each of the possible four spin-1/2 combinations for the input state, the following four Bell-states may be generated:
\index{Bell states}
\index{Spin 1/2}
\begin{equation}
	\begin{array}{c|c}
		\hline \hline \text { In } & \text { Out } \\
		\hline|00\rangle & (|00\rangle+|11\rangle) / \sqrt{2} \equiv\left|\beta_{00}\right\rangle \\
		|01\rangle & (|01\rangle+|10\rangle) / \sqrt{2} \equiv\left|\beta_{01}\right\rangle \\
		|10\rangle & (00\rangle-|11\rangle) / \sqrt{2} \equiv\left|\beta_{10}\right\rangle \\
		|11\rangle & (01\rangle-|10\rangle) / \sqrt{2} \equiv\left|\beta_{11}\right\rangle \\
		\hline \hline
	\end{array}
\end{equation}

The bell states with their commonly used symbolic representations \cite{nielsen_chuang_2010}:
\begin{equation}
	\begin{array}{ll}
		\left|\Phi^{+}\right\rangle=\frac{1}{\sqrt{2}}(|00\rangle+|11\rangle) & \left|\Phi^{-}\right\rangle=\frac{1}{\sqrt{2}}(|00\rangle-|11\rangle) \\
		\left|\Psi^{+}\right\rangle=\frac{1}{\sqrt{2}}(|01\rangle+|10\rangle) & \left|\Psi^{-}\right\rangle=\frac{1}{\sqrt{2}}(|01\rangle-|10\rangle)
	\end{array}
\end{equation}

It should be emphasised to the student that if the spin orientation of the first particle is \textit{\textbf{measured}}, the spin orientation of the second particle is \textbf{\textit{also immediately known,}} regardless of how far away it is. Consequently, such measurements \textbf{\textit{eliminate entanglement}}, forcing both particles into well-defined spin states.

\index{Paper-and-pencil}
\index{Quantum circuit}
\textbf{\raisebox{-0.2em}{\textcolor{orangebdark}{\rule{0.8em}{0.8em}}}  Paper-and-pencil computation:}\\
In this study, we outline the steps of the Bell-state generation circuit, specifically $\Psi_{1} - \Psi_{3}$ shown in Figure \ref{fig:output_matrix}, and discuss the impact of each unitary operation on the circuit's time evolution. We will also reiterate these steps.
\label{Unitary}
\begin{quantikz}[slice all,remove end slices=1,slice
titles=$\Psi$ \col,slice style=blue,slice label style
={inner sep=1pt,anchor=south west,rotate=30}]
\lstick{$\ket{0}$} &    \qw & \gate{H}  & \ctrl{1}  & \meter{} &Control  \\
\lstick{$\ket{0}$} &   \qw & \qw & \targ{0}  & \meter{} &Target
\end{quantikz}
\index{Control}
\index{CNOT}
\index{Target}
\begin{enumerate}[leftmargin=5.5mm]
	\item At \textbf{step $\Psi_{1}$} the circuit is initialized with $\left|\psi_1\right\rangle=|0\rangle_C \otimes|0\rangle_T=|0,0\rangle$ in order to generate the Bell-state $\frac{1}{\sqrt{2}}(|00\rangle+|11\rangle)$.
	\item At \textbf{step $\Psi_{2}$}, following the application of the unitary Hadamard gate the result is:\\
$\left|\psi_2\right\rangle=\left(H \otimes I_2\right)\left|\psi_1\right\rangle=\left(H \otimes I_2\right)\left(|0\rangle_{C O N T R O L} \otimes|0\rangle_{T A R G E T}\right)=H|0\rangle_C \otimes|0\rangle_T=\frac{1}{\sqrt{2}}(|0\rangle+|1\rangle) \otimes|0\rangle$

which finally equals $\frac{1}{\sqrt{2}}(|00\rangle+|10\rangle)$.
	\item At \textbf{step $\Psi_{3}$} the CNOT gate is applied as follows $\left|\psi_3\right\rangle={CNOT}\left|\psi_2\right\rangle$ resulting in $=\frac{1}{\sqrt{2}} {CNOT}(|00\rangle+|10\rangle)=\frac{1}{\sqrt{2}}(|00\rangle+|11\rangle)$ which finally equals $=\frac{1}{\sqrt{2}}\left(\begin{array}{l}1 \\ 0 \\ 0 \\ 1\end{array}\right)$.

\end{enumerate}

The summery of the steps is presented in \eqref{bell-steps}:
\begin{equation}
	\begin{array}{|c|c|}
		\hline \text { Stage } & \text { Quantum state } \\
		\hline 0 & |0\rangle \otimes|0\rangle \\
		\hline 1 & \frac{|0\rangle+|1\rangle}{\sqrt{2}} \otimes|0\rangle \\
		\hline 2 & \frac{|0\rangle \otimes|0\rangle+|1\rangle \otimes|1\rangle}{\sqrt{2}} \\
		\hline
	\end{array}
\label{bell-steps}
\end{equation}

\index{Quantum circuit}
The exact opposite of the Bell-state creation process is the Bell-state \textit{measurement} circuit depicted in \ref{fig:bell:rev}. For instance:
\index{Bell-state measuerment}
\index{Bell-state measuerment circuit}
\begin{figure}[H]
	\centering
	(a)
		\begin{quantikz}[slice all,remove end slices=1,slice
		titles=$\Psi$ \col,slice style=blue,slice label style
		={inner sep=1pt,anchor=south west,rotate=30}]
		&  \ctrl{1} & \gate{H} & \meter{} && \\
		& \targ{0} & \qw  & \meter{} &&
	\end{quantikz}

(b)
\begin{equation}
	\left(\begin{array}{c}
		0 \\
		0.707 \\
		-0.707 \\
		0
	\end{array}\right)=\left(\begin{array}{l}
		0 \\
		0 \\
		0 \\
		1
	\end{array}\right)
\end{equation}
	\caption{(a) Bell-state detection circuit (b) the result of applying the circuit to the Bell-state.)}
	\label{fig:bell:rev}
\end{figure}
\index{Paper-and-pencil}
\textbf{\raisebox{-0.2em}{\textcolor{orangebdark}{\rule{0.8em}{0.8em}}}  Paper-and-pencil computation:} All the Bell states are pure quantum states since using the density matrix it may be shown  that $(|\Psi\rangle\langle\Psi|)^2=|\Psi\rangle\langle\Psi|$:
\index{Quantum circuit}
\begin{equation}
	\Psi \cdot \Psi^{\mathrm{T}}=\left(\begin{array}{cccc}
		0 & 0 & 0 & 0 \\
		0 & 0.5 & -0.5 & 0 \\
		0 & -0.5 & 0.5 & 0 \\
		0 & 0 & 0 & 0
	\end{array}\right) =(\Psi \cdot \Psi^{\mathrm{T}})^2
\label{density:squared}
\end{equation}
The Quantum proof of an entangled superposition is the presence of \textit{off-diagonal entries} in the Bell-state density matrix operator in E.g\eqref{density:squared}.

\subsection {Greenberger Horne Zeilinger (GHZ) nonlocality state generation}
\label{sub:sup:had2}
\index{GHZ}
\index{Greenberger Horne Zeilinger}
\index{Three qubit entanglement}
While the Bell-states involve only two spin-1/2 particles, the Greenberger-Horne-Zeilinger (GHZ) state is a special type of quantum entanglement involving at least three such particles. In 1989, D. Greenberger, M.A. Horne, and Anton Zeilinger were the first to investigate quantum states and reveal their nonclassical features. For quantum systems with  $n=3$ qubits, there are two main types of entanglement, which are exemplified by the GHZ state. First, the GHZ which is a generalization of the Bell-state $\left|\Phi^{+}\right\rangle=$ $1 / \sqrt{2}(|00\rangle+|11\rangle)$ increasing from $n=2$ to $n=3$ qubits \eqref{ghz01}:

\begin{equation}
	|\mathrm{GHZ}\rangle=\frac{1}{\sqrt{2}}(|000\rangle+|111\rangle)
\label{ghz01}
\end{equation}
The 3-qubit state, cannot be written as a tensor product of 1-qubit states.
And the W state, is the second type \eqref{ghz02} which is a generalization of the Bell-state $\left|\Psi^{+}\right\rangle=1 / \sqrt{2}(|01\rangle+|10\rangle)$ increasing from $n=2$ to $n=3$ qubits.:
\begin{equation}
	|\mathrm{W}\rangle=\frac{1}{\sqrt{3}}(|001\rangle+|010\rangle+|100\rangle)
	\label{ghz02}
\end{equation}
Generally for N=n qubits:
\begin{equation}
	\left|\psi_{\mathrm{GHZ}}\right\rangle=\frac{1}{\sqrt{2}}\left(|0\rangle^{\otimes n}+|1\rangle^{\otimes n}\right)
\end{equation}

\index{Quantum circuit}
Here is the 3-qubit GHZ generation circuit \cite{nielsen_chuang_2010}:
\begin{figure}[H]
	\centering
	(a)
	\begin{lstlisting}[caption={},language=Ada]
		cr = chain(
				3,
				repeat(H, 1:3),
				control(2, 1=>X),
				control(3, 1=>X),
				repeat(H, 2:3),
		)
	\end{lstlisting}
	(b)
		\begin{figure}[H]
		\centering
		\includegraphics[totalheight=3cm]{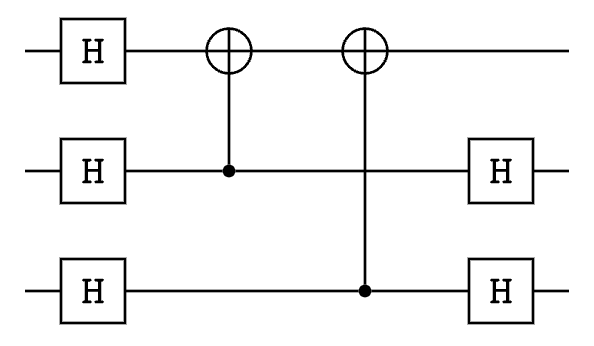}
	\end{figure}
	(c)
	\begin{lstlisting}[caption={},language=Ada]
	$\psi 1=$ ArrayReg(bit"000") $\mid>$ normalize!
	$\psi 1=(\psi 1 \mid>\mathrm{cr})$
	$@show$ $((\psi 1$. state $))$
	> julia
		Matrix{ComplexF64}:
		0.7071067811865472 + 0.0im
		0.7071067811865472 + 0.0im
	\end{lstlisting}
	\caption{(a) Yao.jl code for generating the GHZ states  (b) the circuit diagram (c) the result of applying the circuit to the GHZ state.)}
	\label{fig:ghz}
\end{figure}
\index{Paper-and-pencil}
\textbf{\raisebox{-0.2em}{\textcolor{orangebdark}{\rule{0.8em}{0.8em}}}  Paper-and-pencil computation:} In this study, we outline the steps of the GHZ state generation circuit, specifically $\Psi_{1} - \Psi_{4}$ shown in Figure \ref{fig:ghz:gate:gen}, and discuss the impact of each unitary operation on the circuit's time evolution.
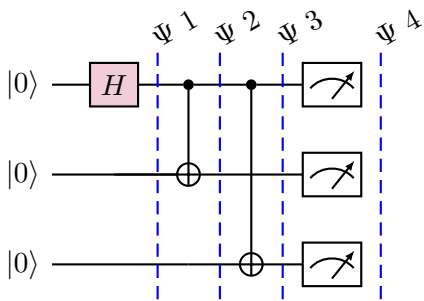
\begin{figure}[H]
	\begin{quantikz}[slice all,remove end slices=1,slice
		titles=$\Psi$ \col,slice style=blue,slice label style
		={inner sep=1pt,anchor=south west,rotate=30}]
		\lstick{$\ket{0}$} &   \gate{H}  & \ctrl{1} & \ctrl{2}  & \meter{} && \\
		\lstick{$\ket{0}$} &   \qw & \targ{0} \qw & \qw &  \meter{} && \\
		\lstick{$\ket{0}$} &   \qw & \qw &  \targ{0}  & \meter{} &&
	\end{quantikz}
	\caption{A quantum circuit that creates the Greenberger-Horne-Zeilinger state among multiple qubits using a quantum circuit model. A step-by-step examination of the GHZ-type entanglement state generating circuit to highlight the influence of the gates at each level as the circuit advances.}
	\label{fig:ghz:gate:gen}
\end{figure}

\begin{enumerate}[leftmargin=5.5mm]
	\item At \textbf{step $\Psi_{1}$} the circuit is initialized with zeros $\left|\psi_1\right\rangle=|0\rangle|0\rangle|0\rangle$  in order to generate the first GHZ state type.
	\item At \textbf{step $\Psi_{2}$}, following the application of the unitary Hadamard gate the result is $\left|\psi_2\right\rangle=\left(H|0\rangle\right) \otimes|0\rangle|0\rangle=\frac{1}{\sqrt{2}}\left(|0\rangle+|1\rangle\right)|0\rangle|0\rangle$
     \item At \textbf{step $\Psi_{3}$} the CNOT gate is applied as follows $\left|\psi_3\right\rangle={CNOT}\left|\psi_2\right\rangle$\\
     $\left|\psi_3\right\rangle=\left(CNOT\right)_{1 2} \frac{1}{\sqrt{2}}\left(|0\rangle_1+|1\rangle_1\right)|0\rangle_2 \otimes|0\rangle_3$ which equals $=\frac{1}{\sqrt{2}}\left(|0\rangle_1|0\rangle_2+|1\rangle_1\right)|1\rangle_2 \otimes|0\rangle_3$
    \item At \textbf{step $\Psi_{4}$} the CNOT gate is applied once again producing $=\frac{1}{\sqrt{2}}(|000\rangle+|111\rangle)=\frac{1}{\sqrt{2}}\left(\begin{array}{l}1 \\ 0 \\ 0 \\ 0 \\ 0 \\ 0 \\ 0 \\ 1\end{array}\right)$

\end{enumerate}

\subsubsection{Local realism}
No discussion on entanglement can be complete without mentioning the concept of local realism. A central theme In local realism is that objects have properties independent of the action of measurement \cite{Ghirardi2005, Greenberger7, Mermin1998}, and a measurement at one location on a first particle has no effect on a measurement of a second particle at a distant location, regardless of the fact that both particles were formed in the same event. Conceptually, local realism holds that spin-1/2 particles contain attributes or instruction sets that determine the outcomes of subsequent measurements. In this school of thought, particles are in a state that can be precisely described even before any measurements are conducted.
\index{Spin-1/2}
\index{Local realism}
\index{EPR}
\index{Maximally entangled}
\index{Hardy'd paradox}
Hardy \cite{Ghirardi2005} established in his fundamental study on local realism that for any entangled, but not maximally entangled, states of two spin-1/2 particles, non-locality may be demonstrated without the use of inequalities. Particularly, he demonstrated that "local reality" is inconsistent with quantum predictions by employing classic EPR-counterfactual arguments.  For the interested reader, formal proofs of these arguments are discussed in detail in  \cite{Frauchiger2018, Aharonov2002, Ghirardi2005, Greenberger7, Mermin1998} and in Hardy's test \cite{Ghirardi2005, Greenberger7,Aharonov2002}, an inequality-free  variant of Bell's theorem \cite{Dong2020, Mermin1998} which invalidates locality.

%
%

\section{Conclusions}

Quantum computing has the potential to transform a number of fields, including medicine, finance, and security. However, it can be challenging for students, particularly those studying computer science, to fully grasp the concept of entanglement, which is a key component of quantum computing, due to its complexity and abstract nature.

In the past, courses on quantum computing have only been offered by physics departments, but in order to adequately prepare computer scientists for careers in the quantum industry, new training programs need to be developed in engineering and computer science departments in collaboration with applied physics departments.

One approach to helping students better understand entanglement and quantum computing is to use the programming language Julia, which is specifically designed for scientific computing and allows students to learn through hands-on, interactive experiences. It is important for computer science students to focus on the practical applications and potential uses of entanglement through programming quantum circuits using Python or Julia.

Meanwhile, a more theoretical approach, which focuses on the fundamental principles of entanglement from the perspective of QM and their application to quantum computing, as well as the technical aspects of quantum algorithms and hardware, may be more suitable for students of physics. It is important to find a balance between these approaches and tailor them to the level of familiarity and interest of the audience.

Unfortunately, \textbf{\textit{it seems unlikely that Julia will become widely adopted in the field of quantum computing,}} as most major vendors have chosen to use Python as their primary programming language for developing their quantum platforms. This follows a similar pattern seen with artificial intelligence libraries such as PyTorch and TensorFlow.

\section{Acknowledgements.}
We thank the authors of Yao.il (\cite{Luo2020Yao}) Luo, Xiu-Zhe and Liu, Jin-Guo and Zhang, Pan and Wang, Lei for their responses on the Yao.jl discussion group.

\bibliographystyle{alphaurl}
\onecolumn
\listoffigures
\listoftables
\bibliography{refs}

\renewcommand\indexname{Index}
\printindex
\end{document}